\documentclass[structabstract]{aa}  
\usepackage{graphicx}
\usepackage{txfonts}
\usepackage{natbib}
\begin{document}
   \title{The Coalsack near and far}


   \author{H.~Beuther
          \inst{1}
          \and
          J.~Kainulainen
          \inst{1}
          \and
          Th.~Henning
           \inst{1}
          \and
          R.~Plume
           \inst{2}
          \and
          F.~Heitsch
          \inst{3}
           }
   \institute{$^1$ Max-Planck-Institute for Astronomy, K\"onigstuhl 17,
              69117 Heidelberg, Germany, \email{beuther@mpia.de}\\
$^2$ Department of Physics and Astronomy, University of Calgary, Calgary, Canada\\
             $^3$ Department of Physics \& Astronomy, University of North Carolina at Chapel Hill, CB 3255, Chapel Hill, NC 27599-3255, USA}



\abstract
{The large Coalsack dark cloud is one of the most prominent southern
  starless clouds, which is even visible to the naked eye. Furthermore,
  it is one of the rare molecular clouds without clear signs of star
  formation.}
{We investigate the dynamical properties of the gas within the
  Coalsack.}
{The two highest extinction regions were mapped with the APEX
  telescope in $^{13}$CO(2--1) comprising a region of $\sim$1\,square
  degree.}
{In addition to the well-known, nearby gas component around
  $-4$\,km\,s$^{-1}$, we identified additional molecular gas
  components -- in particular a second extended molecular cloud at a
  velocity of $\sim -30$\,km\,s$^{-1}$ and an estimated distance of
  $\sim$3.1\,kpc -- that dominate the column density and visual
  extinction distributions in the northeastern part of the Coalsack.
  Although comprising $\sim$2600\,M$_{\odot}$, the mass of this
  distant cloud is distributed over an extent of $\sim$73\,pc, much
  larger than typical high-mass infrared dark clouds. Its filamentary
  structure is consistent with a compressible gaseous self-gravitating
  cylinder, and its low mass per length indicates that it may be
  stable against gravitational collapse.  We find barely any
  mid-infrared emission in archival MSX data, which is indicative of
  almost no star-formation activity in the near and far cloud
  complexes. The nearby clouds have narrow, almost thermal velocity
  dispersions with median values between 0.2 and 0.4\,km\,s$^{-1}$,
  which is also consistent with low star-formation activity. Only
  Tapia's Globule 2 exhibits a velocity dispersion increase toward the
  extinction peak and peak-velocity gradients over the core, which is 
  indicative of a state of elevated dynamical properties.}
{The Coalsack is not one single coherent structure, but consists of
  several cloud complexes nearby as well as at several kpc distance.
  All studied clouds appear as starless low-turbulence regions that
  may not even collapse in the future. Only one globule exhibits more
  dynamical signatures and is a good candidate for present/future star
  formation.}  
   \maketitle

\section{Introduction}
\label{intro}

\begin{figure}[htb]
\includegraphics[angle=-90,width=90mm]{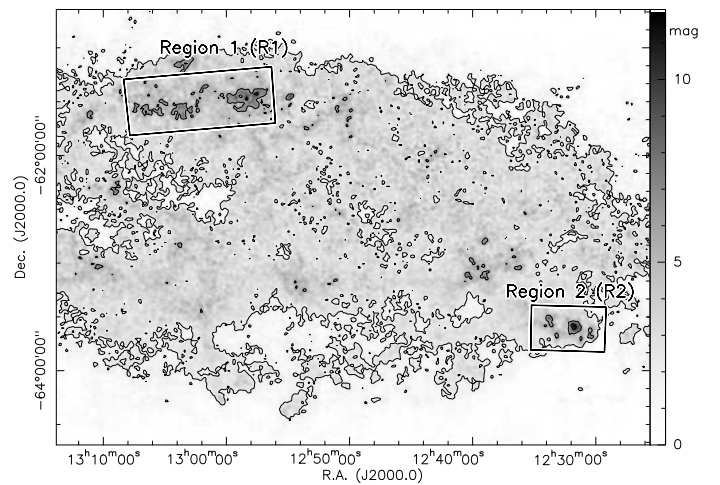}
\caption{2MASS extinction map of the whole Coalsack by
  \citet{kainulainen2009}. The contour levels are at 2, 6 and 10\,mag
  extinction. The two boxes outline the two regions mapped in
  $^{13}$CO(2--1) and C$^{18}$O(2--1) for this work.}
\label{extinction}
\end{figure}

The Coalsack is one of the closest and most widely-recognized dark
clouds in the southern hemisphere (between Galactic longitudes
$305\geq l \geq 300$\,deg) because it is visible to the bare eye
(e.g., \citealt{nyman2008}).  \citet{kainulainen2009} recently
produced extinction maps based on 2MASS near-infrared data of all
supposedly nearby molecular clouds including the Coalsack. Figure
\ref{extinction} shows the large-scale extinction ($A_v$) map of this
region. While the visual extinction through the Coalsack is generally
low, there are a few regions where the extinction $A_v$ rises above 6
and even 10\,mag. Although the extinction maps set significant
constraints on the morphology and column density distributions,
allowing us to derive results about cloud structure and stability as
well as potential column density star-formation thresholds
(\citealt{kainulainen2009,kainulainen2011}), these data lack any
kinematical information about the cloud. Furthermore, this cloud is
one of the rare cases of a nearby molecular cloud devoid of any
star-formation signatures (e.g., \citealt{nyman2008,kainulainen2009}).
This is one of the reasons why it has not received as much attention
in the past as typical star-forming clouds like Taurus of $\rho$
Ophiuchus. Despite this, the Coalsack therefore constitutes an
excellent laboratory to study pristine and undisturbed molecular
clouds.

Earlier spectral-line mapping on the Coalsack was conducted in
$^{12}$CO by \citet{nyman1989} at a spatial resolution of 8.8$'$, and
in $^{13}$CO by \citet{kato1999} at a resolution of 2.7$'$. Thus, even
basic dynamical measures (e.g., central velocities, dispersions, etc.)
were limited to spatial scales larger than what is expected for
molecular clumps or cores (or equally limited in dynamical
range/velocity resolution).  The typical velocities of the region are
between -6 and 0\,km\,s$^{-1}$ (e.g., \citealt{nyman1989}), and the
distance estimates vary between 100 and 200\,pc (e.g.,
\citealt{franco1995,corradi1997,knude1998,nyman2008}).
\citet{nyman1989} reported the presence of an additional velocity
component at $-35$\,km\,s$^{-1}$, which they attribute to a background
cloud associated with the Sagittarius-Carina arm that is unlikely to
be responsible for any of the extinction features.  Some individual
fields of the Coalsack have been mapped in tracers of higher density
by some authors, mainly concentrating on the region known as ``Tapia's
Globule 2'', which presumably is one of the few dense cores in the
region (e.g., \citealt{tapia1973,lada2004,rathborne2009b}).
Furthermore, \citet{fontani2005}, \citet{beltran2006} and
\citet{miettinen2010} studied a smaller sub-filament just adjacent to
the eastern edge of our northeastern field. These studies, however,
cover only a minuscule portion of the complex and therefore are not very
informative for the dynamics of the complex itself.

\begin{figure}[htb]
\includegraphics[width=90mm]{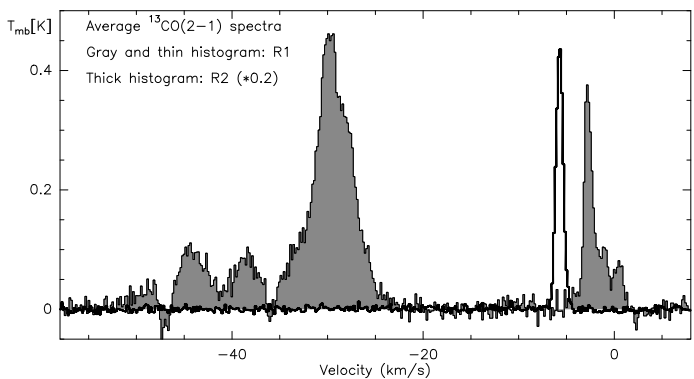}
\caption{Average $^{13}$CO(2--1) spectra toward the northeastern
  region R1 (gray with this histogram) and the southwestern region R2
  (thick histogram multiplied by 0.2 to be on the same scale).}
\label{aver}
\end{figure}

To fill this missing link was one of the motivations mapping parts of
the Coalsack in spectral line emission. For this purpose, we selected
the two major regions with significant fractions of cloud structure
above $\sim$6\,mag extinction, marked by the two boxes in Figure
\ref{extinction}. The southwestern region also comprises ``Tapia's
Globule 2''. These two regions cover in total about 1 square degree
and were mapped with APEX in $^{13}$CO(2--1) and C$^{18}$O(2--1) at a
spatial resolution of $\sim 27.5''$.

\section{Observations and data} 
\label{obs}

\begin{figure*}[htb!]
\includegraphics[width=184mm]{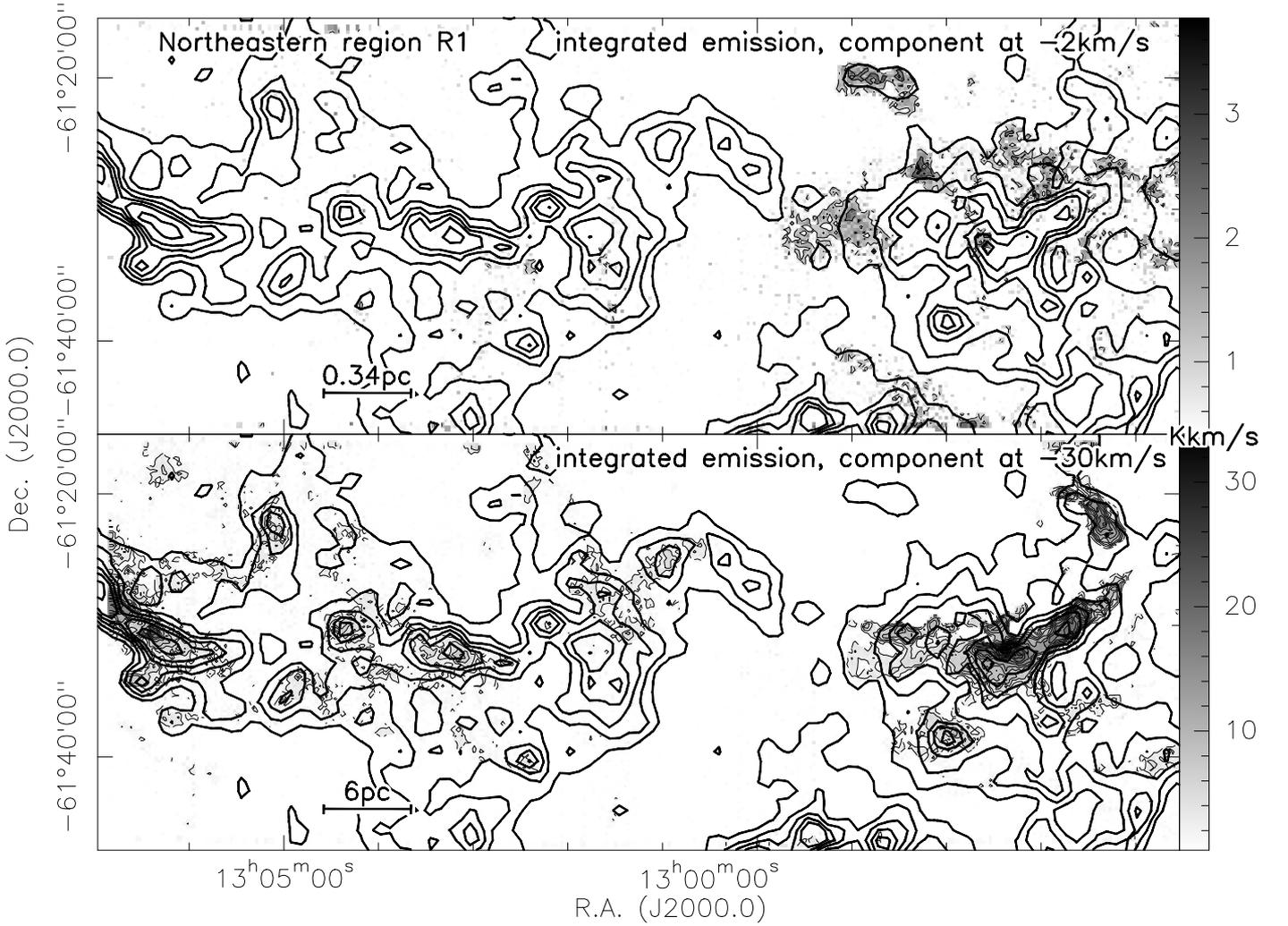}
\caption{Compilation of the integrated emission in grayscale (with
  thin contours) from the -2\,km\,s$^{-1}$ and -30\,km\,s$^{-1}$
  components in $^{13}$CO(2--1) toward the northeastern region R1 in
  the top and bottom panels, respectively. The thick contours outline
  the extinction from \citet{kainulainen2009} in 1\,mag steps between
  $A_v$ of 5 and 9\,mags plus additional contours of 11 and 13\,mags.
  The integration regimes for the top and bottom panels are [-4,2] and
  [-35,-24]\,km\,s$^{-1}$, respectively.  The thin contours for the
  -4\,km\,s$^{-1}$ map start at 1\,K\,km\,s$^{-1}$ and continue in
  1\,K\,km\,s$^{-1}$ steps, whereas for the -30\,km\,s$^{-1}$ component
  the contours start at 2\,K\,km\,s$^{-1}$ and continue in
  2\,K\,km\,s$^{-1}$ steps. A linear scale-bar is shown in both
  panels}
\label{overlay_r1}
\end{figure*}

The C$^{18}$O(2--1) and $^{13}$CO(2--1) data at 219.560\,GHz and
220.399\,GHz were observed simultaneously with the Atacama Pathfinder
Experiment (APEX\footnote{This publication is based on data acquired
  with the Atacama Pathfinder Experiment (APEX). APEX is a
  collaboration between the Max-Planck-Institut fur Radioastronomie,
  the European Southern Observatory, and the Onsala Space
  Observatory.}, \citealt{guesten2006}) between April and June 2010 in
the 1\,mm band for the two regions R1 and R2 marked in
Fig.~\ref{extinction} in on-the-fly mode. The APEX1 receiver of the
SHeFI receiver family has receiver temperatures of $\sim$130\,K at
220\,GHz \citep{vassilev2008}, and the average system temperatures
during the observations were $\sim$220\,K.  Two
Fast-Fourier-Transform-Spectrometer (FFTS, \citealt{klein2006}) were
connected, covering $\sim$2\,GHz bandwidth between 219 and 221\,GHz
with a spectral resolution of $\sim$0.17\,km\,s$^{-1}$. The data were
converted to main-beam brightness temperatures $T_{\rm{mb}}$ with
forward and beam efficiencies at 220\,GHz of 0.97 and 0.82,
respectively \citep{vassilev2008}. The average 1$\sigma$ rms in the
maps of emission-free channels is $\sim$0.75\,K. Moment maps were
conducted clipping all pixels below a 2\,K threshold. The selected
OFF-positions were at the edge of the Coalsack and were free of
apparent CO emission (R.A. (J2000) 12:57:20.514 Dec. (J2000)
-60:59:43.42 for R1 and R.A. (J2000) 12:31:51.0 Dec. (J2000)
-63:57:56.9 for R2). The FWHM of APEX at the given frequencies is
$\sim 27.5''$.

The near-infrared extinction map is taken from
\citet{kainulainen2009}. The spatial resolution of this map is $\sim
90''$.

\section{Results}

\subsection{The Coalsack: Near and far components}
\label{results}

The $^{13}$CO(2--1) emission in the northeastern region R1 exhibits
several velocity components (see average spectrum in Fig.~\ref{aver}),
most prominently the well-known Coalsack component at $\sim
-2$\,km\,s$^{-1}$ as well as a second even stronger component at $\sim
-30$\,km\,s$^{-1}$. Furthermore, there are additional components at
more negative velocities, one centered at $\sim
-38$\,km\,s$^{-1}$ and one at $\sim -44$\,km\,s$^{-1}$. The
south-eastern region R2 exhibits only one velocity component at $\sim
-4$\,km\,s$^{-1}$. The little velocity offsets between the $\sim
-2$\,km\,s$^{-1}$ and $\sim -4$\,km\,s$^{-1}$ components of regions R1
and R2 imply a small large-scale velocity gradient over the entire
Coalsack complex. How do these components now correspond to the
extinction in the Coalsack?

\begin{figure}[htb!]
\includegraphics[width=90mm]{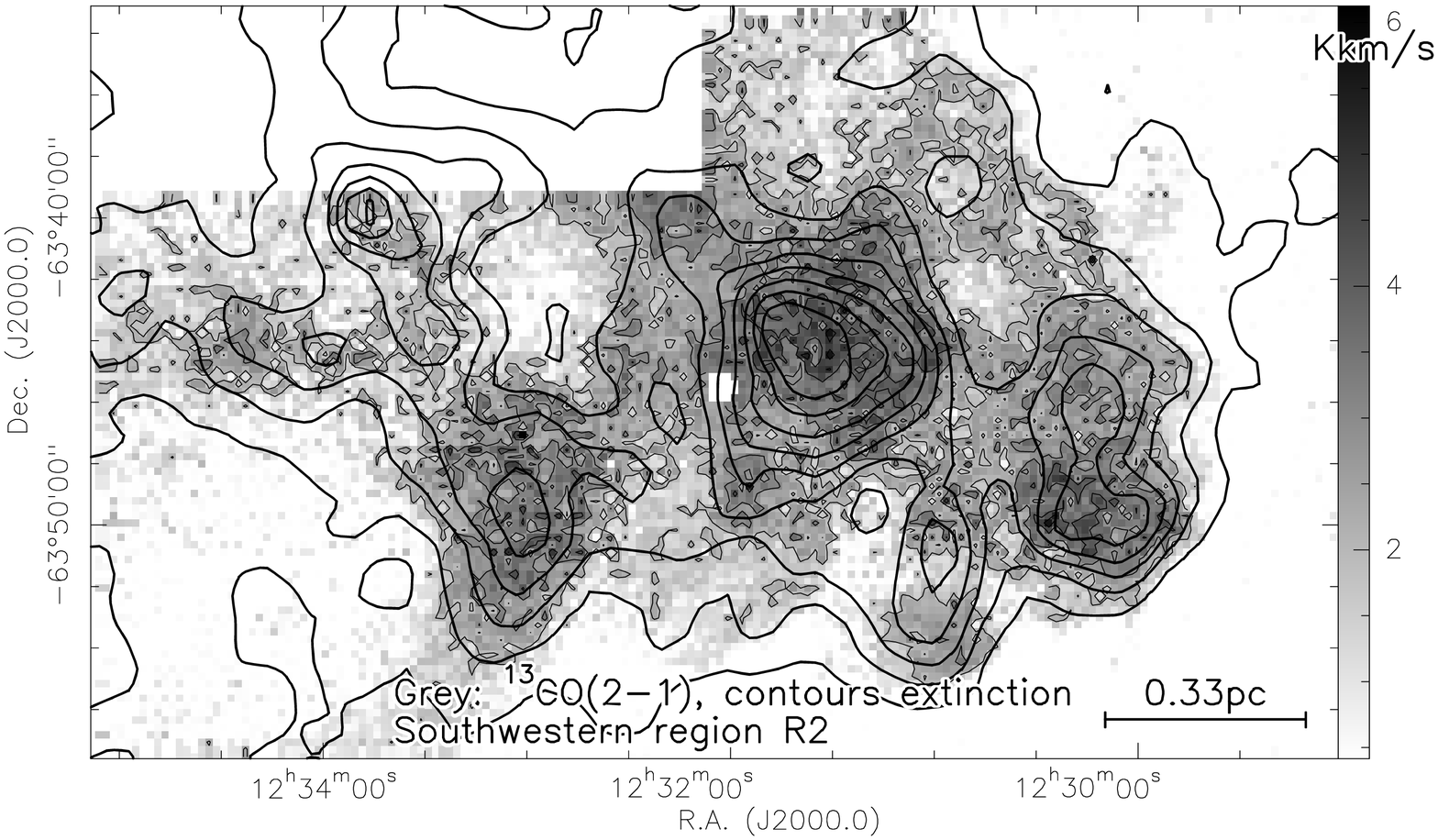}
\caption{The grayscale with thin contours shows the integrated
  $^{13}$CO(2--1) emission toward the southwestern region R2. Contour
  levels start at 2\,K\,km\,s$^{-1}$ and continue in
  1\,K\,km\,s$^{-1}$ steps. The thick contours outline the extinction
  from \citet{kainulainen2009} in 1\,mag steps starting at an $A_v$ of
  2\,mags. A linear scale-bar is shown in the bottom-right corner.}
\label{overlay_r2}
\end{figure}

\begin{figure}[htb]
\includegraphics[width=90mm]{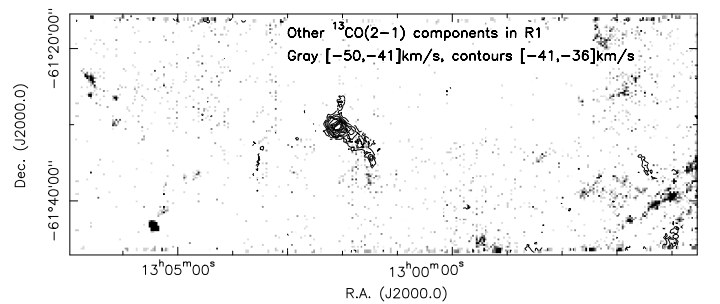}
\caption{Integrated $^{13}$CO(2--1) images in the northern R1 region
  of the two other velocity components seen in Fig.~\ref{aver}. The
  grayscale shows the integration regime [-41,-36]\,km\,s$^{-1}$, and
  the contours present the velocity regime [-50,-41]\,km\,s$^{-1}$.}
\label{r1_13co_other}
\end{figure}

Figures \ref{overlay_r1} and \ref{overlay_r2} present integrated
$^{13}$CO(2--1) maps of the strongest emission features toward the
northeastern region R1 and the southwestern region R2, respectively.
While the $\sim -4$\,km\,s$^{-1}$ component in region R2 exhibits the
expected structure, peaking in the direction of the extinction peak of
Tapia's globule 2 (see also \citealt{lada2004,rathborne2009b}), the
region R1 clearly exhibits two spatially very distinct velocity
components around $\sim -2$ and $\sim -30$\,km\,s$^{-1}$.
Furthermore, Figure \ref{r1_13co_other} presents the integrated
$^{13}$CO(2--1) maps of the additional weaker components at $\sim
-38$\,km\,s$^{-1}$ and $\sim -44$\,km\,s$^{-1}$. While we can identify
spatial structures from these two velocity components as well, they
are less prominent and also less extended. Therefore, we omit the
latter two velocity components in the following discussion.

The C$^{18}$O(2--1) emission is comparably weak and we only barely
detect some structures.  While we detect C$^{18}$O(2--1) in the $\sim
-4$\,km\,s$^{-1}$ component toward R2, we only detect the $\sim
-30$\,km\,s$^{-1}$ component toward R1.  Therefore, here we only work
with the $^{13}$CO(2--1) data.

Although \citet{nyman1989} reported spectra with a second velocity
component, their mentioned positions were far offset from the
northeastern region R1.  In addition to this, Figure \ref{overlay_r1}
clearly shows that the extinction is not caused by the nearby
component at the velocity of $\sim -2$\,km\,s$^{-1}$, but is
mainly associated with the second much stronger velocity component
at $\sim -30$\,km\,s$^{-1}$.  This is important because it implies
that the visual dark cloud Coalsack is not one coherent low-mass cloud
as believed so far, but that it is comprised of several clouds at
different distances, all combined to produce the dark features on the
sky. The dark features come primarily from the distant component.

\begin{figure*}[htb!]
\centerline{\includegraphics[width=156mm]{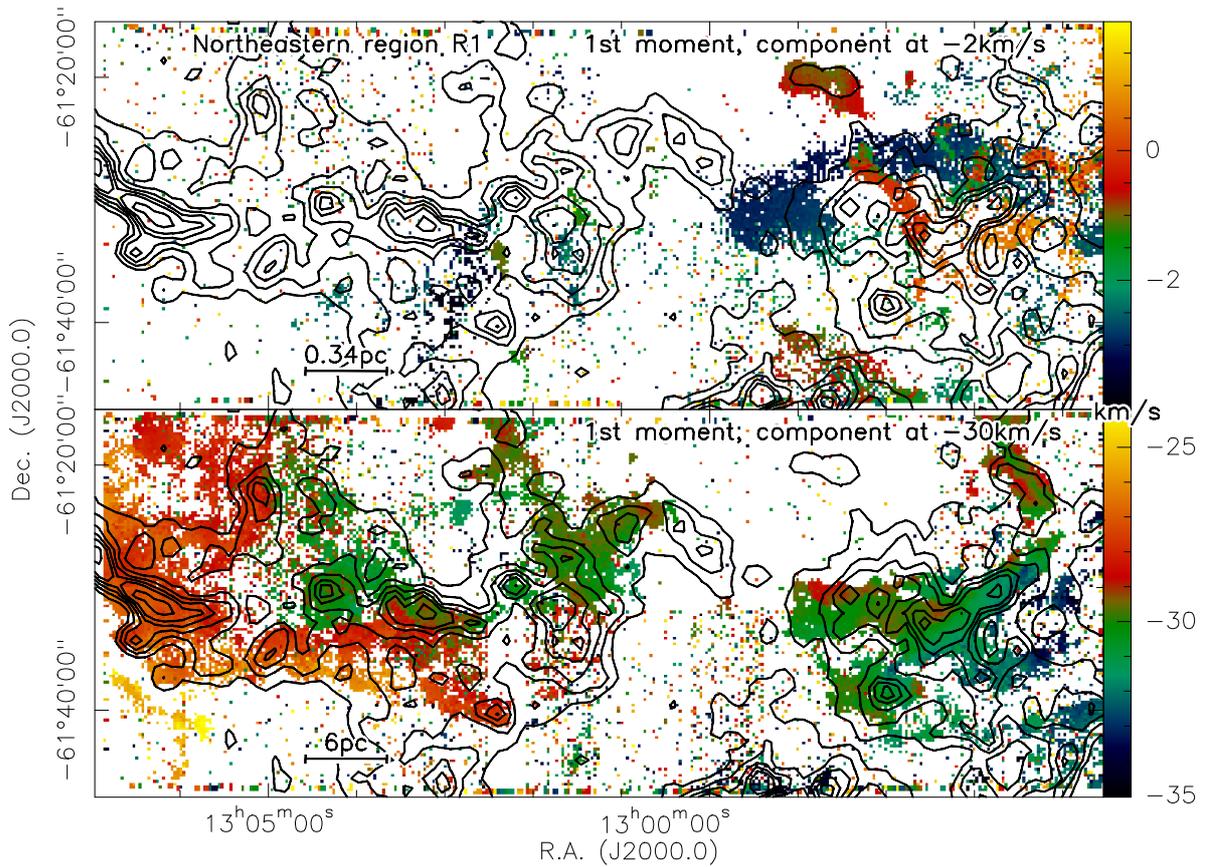}}
\caption{Compilation of $^{13}$CO(2--1) 1st moment maps toward the
  northeastern region R1 as marked in the panels. The top and bottom
  panel show the -2 and -30\,km\,s$^{-1}$, respectively. The thick
  contours outline the extinction from \citet{kainulainen2009} in
  1\,mag steps starting between $A_v$ of 5 and 9\,mags plus additional
  contours at 11 and 13\,mags. Linear scale-bars are shown in the
  bottom-left corner of each panel.}
\label{moment1_r1}
\end{figure*}

Following the rotation curves by either \citet{brand1993} or
\citet{reid2009}, for the $-30$\,km\,s$^{-1}$ component the kinematic
near distances are 2.7 or 3.1\,kpc, respectively. Because we see the
Coalsack in extinction, we can safely take the near distance since at
far distances one would not see any extinction. For comparison,
\citet{fontani2005} derived a distance of 2.44\,kpc for the position
of the IRAS\,13039$-$6108 source from CS data. They used a velocity of
$-26.2$\,km\,s$^{-1}$, consistent with our data at the eastern edge of
the map (see Fig.~\ref{moment1_r1}), as well as the \citet{brand1993}
rotation curve. Because we use an average velocity over the whole
structure for the distance determination (Fig.~\ref{moment1_r1}) and
try out different rotation curves, we will stick to the distance of
3.1\,kpc derived with the most recent rotation curve \citep{reid2009}.
Qualitatively speaking, these results imply that not only a small
sub-filament around the IRAS source, but the whole northeastern high
extinction region R1 is not, as usually assumed, a low-mass dark cloud
at distances below 200\,pc, but is a more massive molecular cloud
at a distance of several kpc.

\begin{figure}[htb!]
\includegraphics[angle=-90,width=90mm]{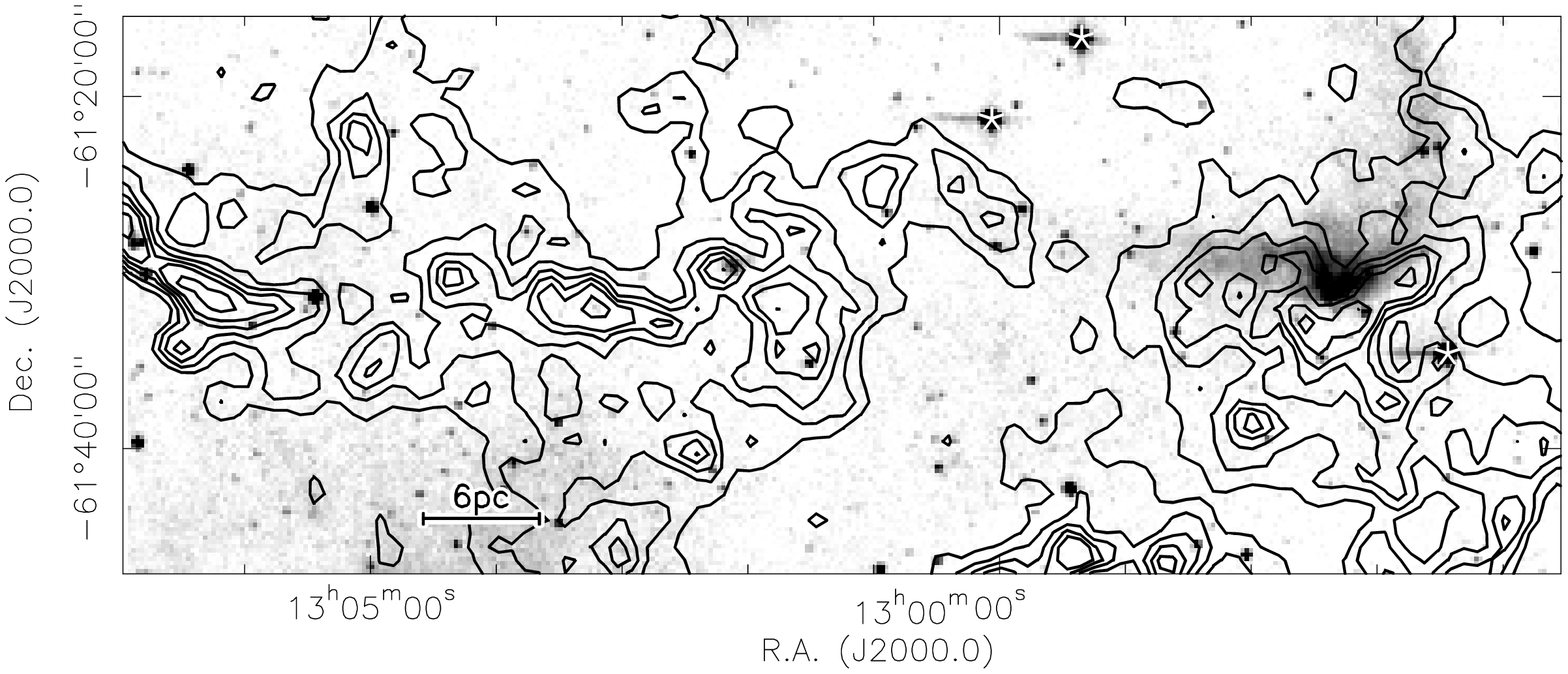}
\includegraphics[angle=-90,width=90mm]{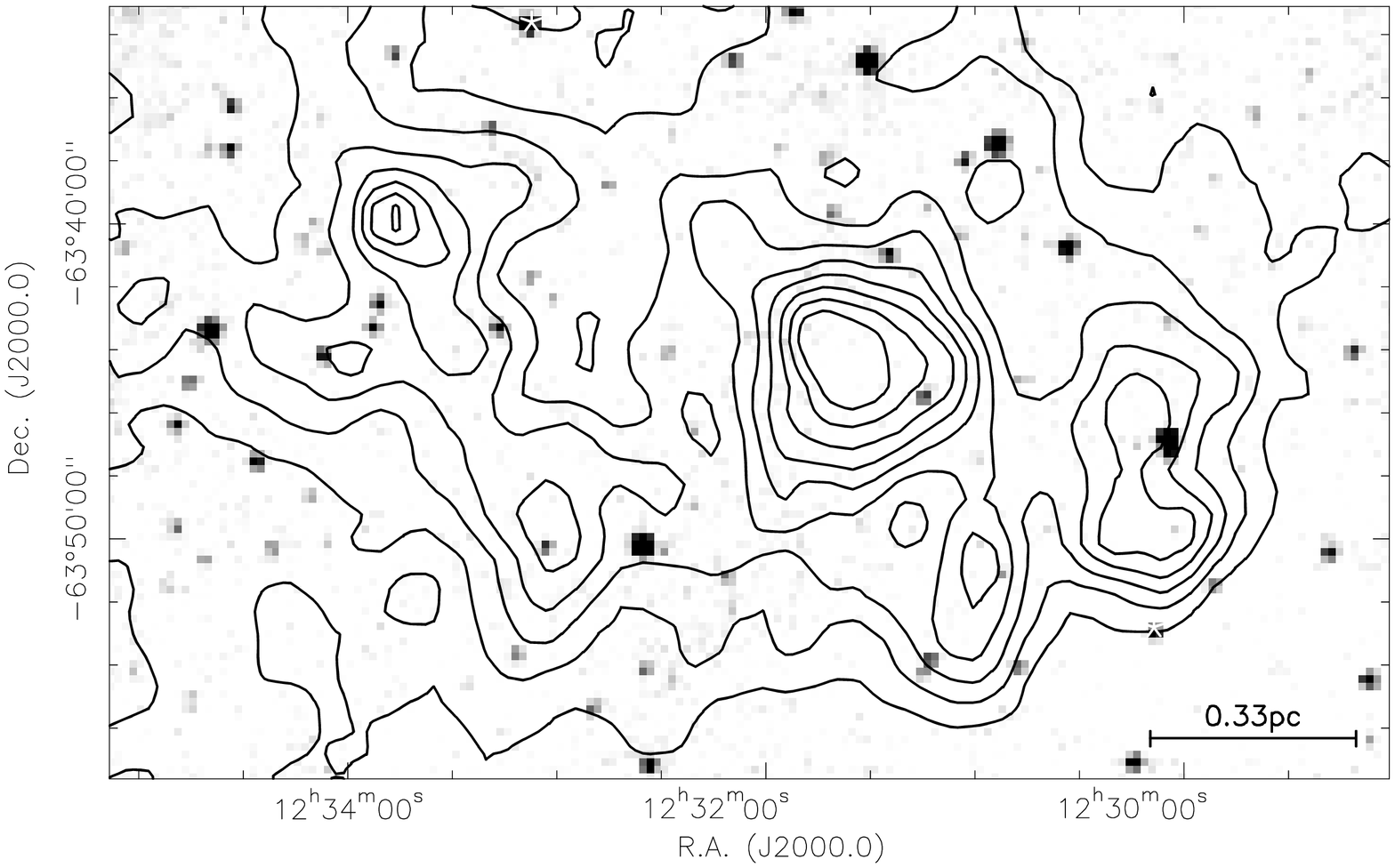}
\caption{The grayscaling presents the 8\,$\mu$m data toward the R1
  (top) and R2 (bottom) regions as observed with MSX. The contours
  outline the extinction from \citet{kainulainen2009} for R1 in 1\,mag
  steps between $A_v$ of 5 and 9\,mags plus additional contours at 11
  and 13\,mags, and for R2 in 1\,mag steps starting at an $A_v$ of
  2\,mags. The white stars mark the only sources that were also
  detected with MSX at 21\,$\mu$m.}
\label{overlay_msx}
\end{figure}

Employing the Galactic structure outlined in \citet{reid2009}, our
lines-of-sight pass through the near and far sides of the Galactic
spiral arms Carina-Sagittarius and Crux-Scutum ($\leq$2\,pc and
$\geq$3\,kpc distance, respectively). While the Carina-Sagittarius arm
should accordingly be responsible for the background structures of the
nearby cloud components, likely the Crux-Scutum arm -- in which the
distant cloud component should be embedded -- also provides the
emission against which to produce the observed extinction features.

Assuming optically thin $^{13}$CO(2--1) line emission, we can
calculate the gas masses for the different components following
\citet{rohlfs2006}. The $-$30\,km\,s$^{-1}$ component in region R1 is
assumed to be at a distance of 3.1\,kpc (labeled R1far from now on),
whereas for the -2/-4\,km\,s$^{-1}$ components in R1 and R2 we use a
distance of 175\,pc (and label it R1near in the northeastern region
R1). Because all clouds are extinction features and hence are largely
comprised of cold, mainly starless gas, an average temperature of
15\,K appears reasonable. With these numbers, we derive a mass of
$\sim$2600\,M$_{\odot}$ for the -30\,km\,s$^{-1}$ R1 cloud, whereas the
two nearby -2/-4\,km\,s$^{-1}$ of R1 and R2 only have masses of
$\sim$1.2 and $\sim$10.4\,M$_{\odot}$, respectively.  Hence the
30\,km\,s$^{-1}$ cloud at a distance of 3.1\,kpc is 2-3 orders of
magnitude more massive than the other two components at the more
typical Coalsack distance. However, one should keep in mind that a
$\sim$2600\,M$_{\odot}$ cloud, that is distributed over a size of
$\sim 73$\,pc in east-west direction, cannot be considered as a
typical high-mass star-forming region. The mass is quite smoothly
distributed over a large spatial area. For comparison, cloud masses of
typical low- and high-mass star-forming regions like Taurus or Orion
vary between several thousand to several $10^4$\,M$_{\odot}$ (e.g.,
\citealt{kainulainen2009}).

To check for signatures of ongoing star formation, we investigated the
8 and 21\,$\mu$m images of the MSX satellite \citep{price1995}.
\footnote{Our Coalsack targets are only barely covered by Spitzer
  because they are at relatively high and low galactic longitude, and
  GLIMPSE only covered latitudes $\leq\pm 1$\,deg
  \citep{churchwell2009}}. Figure \ref{overlay_msx} presents overlays
of the 8\,$\mu$m emission with the extinction map. The northern region
R1 shows one bright 8\,$\mu$m source with a bit of extended diffuse
emission in the west. This bright region may be associated with the
extinction peak at an approximate separation of $2.6''$, however, no
21\,$\mu$m emission is detected from that region. While the 8\,$\mu$m
peak is neither clearly associated with the -30\,km\,s$^{-1}$ nor with
the -2\,km\,s$^{-1}$ $^{13}$CO component, the more diffuse 8\,$\mu$m
emission extending from this peak toward the north-west is associated
with near-infrared extinction as well as $^{13}$CO(2--1) emission of
the -30\,km\,s$^{-1}$ component. Therefore, it appears more likely
that the 8\,$\mu$m peak is also associated with the -30\,km\,s$^{-1}$
component.  In general, there is barely any 21\,$\mu$m emission
detected toward both regions, implying no or only a marginal level of
embedded star formation. This is consistent with the general picture
of the Coalsack being largely a starless cloud.

\subsection{Kinematic analysis of the Coalsack clouds}

In addition to the distance and mass estimates, the $^{13}$CO(2--1)
data also allow us to characterize the kinematic properties of the two
-- or more correctly three -- regions. Here, we will focus on the
northeastern regions (R1far and R1near) and the southwestern region
R2 separately.

\subsubsection{The more massive and more distant cloud R1far}
\label{r1far}

Figure \ref{moment1_r1} presents the $^{13}$CO(2--1) first-moment maps
(intensity-weighted peak velocities) of the two northeastern regions
R1far and R1near. Covering an east-west length of $\sim 81'$ (or $\sim
1.35$\,deg) equivalent to a linear extent $\sim$73\,pc, the velocity
gradient across this structure is relatively small with a velocity
difference of only $\sim$8\,km\,s$^{-1}$ (from $\sim -26$ to $\sim
-34$\,km\,s$^{-1}$), corresponding to
$\sim$0.1\,km\,s$^{-1}$\,pc$^{-1}$. The relatively smooth velocity
gradient implies that this is one coherent structure, even the parts
with lower extinction and almost no $^{13}$CO(2--1) detection in the
middle of the filamentary structure.

\begin{figure}[htb!]
\includegraphics[width=90mm]{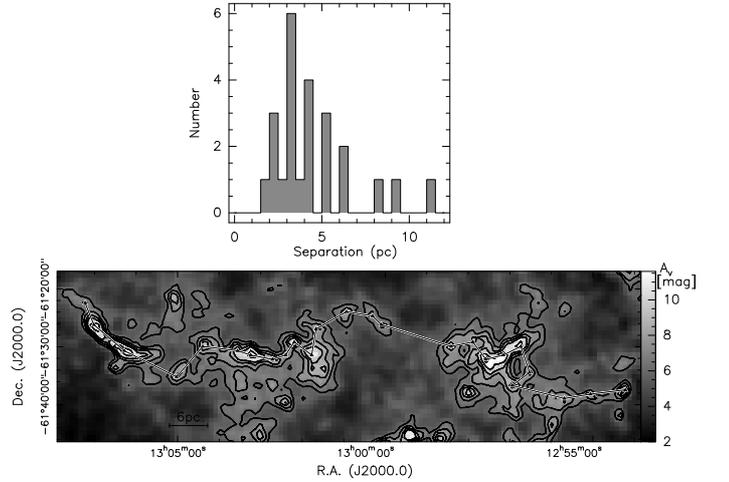}
\caption{Outline of the clump distribution for R1far. The grayscale
  with contours shows the extinction map from \citet{kainulainen2009}
  in 1\,mag steps between $A_v$ of 6 to 9\,mags plus additional
  contours at 11 and 13\,mags. A linear scale-bar is shown at the
  bottom-left. The east-west extent of this extinction map is slightly
  larger than that of the spectral line data. The triangles mark the
  clump peak positions as derived via gaussclump \citep{williams1994},
  and the connecting line visualizes the separations measured between
  the clumps. The histogram above shows the distribution of nearest
  neighbor clump separations as derived from the extinction map.}
\label{separation}
\end{figure}

\begin{figure*}[htb]
\centerline{\includegraphics[width=156mm]{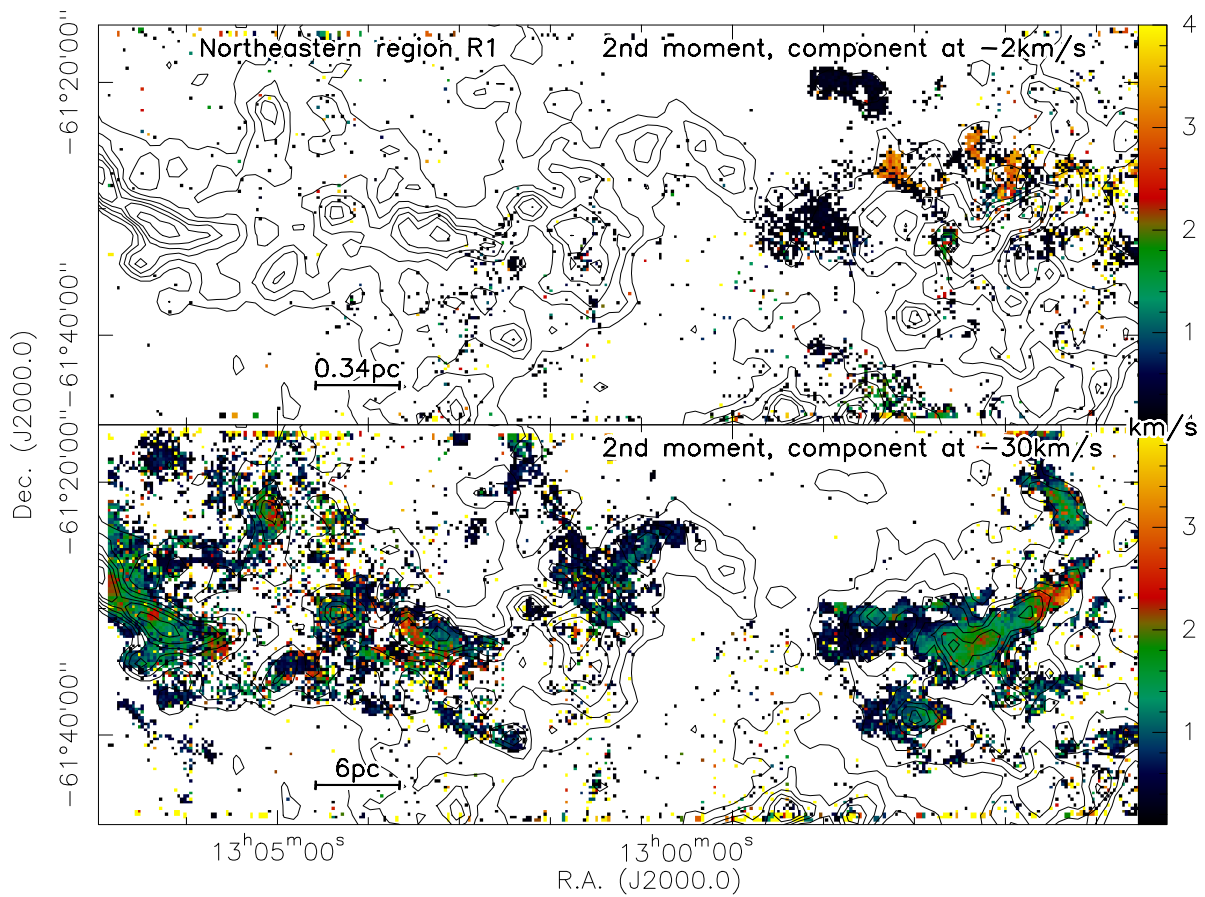}}
\caption{Compilation of $^{13}$CO(2--1) 2nd moment maps toward the
  northeastern region R1 as marked in the panels. The top and bottom
  panel show the -2 and -30\,km\,s$^{-1}$, respectively. The thick
  contours outline the extinction from \citet{kainulainen2009} in
  1\,mag steps between $A_v$ of 5 and 9\,mags plus additional contours
  at 11 and 13\,mags. Linear scale-bars are shown in the bottom-left
  corner of each panel.}
\label{moment2_r1}
\end{figure*}

Inspired by the recent work of \citet{jackson2010} on the large
($\sim$80\,pc length) ``Nessie'' filamentary infrared dark cloud
(IRDC), we investigated whether the $-30$\,km\,s$^{-1}$ component
R1far may also be consistent with a gravitationally bound gaseous
cylinder that may fragment because of fluid instabilities. For that
purpose we used the clumpfind algorithm \citep{williams1994} to
extract clump peak positions from the near-infrared extinction map.
Because we are only interested in the peak positions here, we started
with relatively high 6\,mag contours and continued at 1\,mag levels as
input parameters for clumpfind. Because the filamentary structure is a
bit larger than our $^{13}$CO coverage, we used also a slightly larger
extinction map for the purpose of the filament analysis
(Fig.~\ref{separation}). The projected east-west extent of the
analyzed structure is $\sim$88\,pc. The extracted positions are shown
in Figure \ref{separation}. Based on these peak positions we
calculated the projected linear separations for these peaks assuming a
distance of 3.1\,kpc. One should keep in mind that all observed
separations discussed below have always to be considered as lower
limits because we have no information about the inclination angle of the
filament with the plane of the sky. Figure \ref{separation} also
presents a histogram of the distribution of linear separations, and
the peak of the distribution is at about $\sim$3.5\,pc. While the mean
separation is $\sim$4.6\,pc, the median is at $\sim$4.1\,pc. This
separation is resolved well by the spatial resolution of the
extinction map of $\sim 1.35$\,pc, and random placement of the cores
within the filament would produce a very different distribution with a
mean core separation of $\sim$2\,pc.
The median separation within the R1far filament is
consistent with that derived for the ``Nessie'' IRDC by
\citet{jackson2010}. For more discussion see section \ref{frag_fil}.

Figure \ref{moment2_r1} presents the 2nd moment maps -- or the
intensity-weighted velocity dispersion. While there is a spread in
velocity dispersion between approximately 0.4 and 4\,km\,s$^{-1}$ over
the entire region, most observed line-widths are at the lower end of
the distribution with an average velocity dispersion over the entire
field of $\sim$1.26\,km\,s$^{-1}$ and a median velocity dispersion of
$\sim$0.8\,km\,s$^{-1}$. This relatively narrow average line-width
compared to more evolved star-forming regions confirms the
previous assessment from the near- and mid-infrared data, that the
whole region exhibits low star-formation activity capable of
contributing to the line-width broadening.

\subsubsection{The low-mass nearby cloud R1near}

Similar to R1far, Figures \ref{moment1_r1} and \ref{moment2_r1} also
present the 1st and 2nd moment maps of the -2\,km\,s$^{-1}$ component
in the R1 region. From now on we call that R1near. As already outlined
in section \ref{results}, there is barely any spatial association of
the extinction map with the -2\,km\,s$^{-1}$ $^{13}$CO(2--1) emission.
The velocity spread of the region between R.A. 13:00:00 and the
western edge of the map is only about 4\,km\,s$^{-1}$ between $\sim
-3$ and $\sim 1$\,km\,s. This velocity gradient covers an east-west
extent of $\sim 32'$, corresponding at a distance of 175\,pc to an
approximate linear extent of $\sim 1.6$\,pc.  This results in an
approximate velocity gradient of 2.6\,km\,s$^{-1}$\,pc$^{-1}$. While
this velocity spread appears to be relatively small compared with
the linear velocity gradient one can derive for R1far
($\sim$0.1\,km\,s$^{-1}$\,pc$^{-1}$), it is still comparatively large.
Or the other way round, the velocity gradient for R1far is extremely
small over a much larger spatial extent (see previous section).

The 2nd moment or velocity dispersion distribution for R1near is even
narrower than that for R1far (Fig.~\ref{moment2_r1}). While there are
a few small regions exhibiting velocities $>$1\,km\,s$^{-1}$, the
average and median velocity dispersions for R1near are
0.85\,km\,s$^{-1}$ and 0.2\,km\,s$^{-1}$, respectively. While the
average represents the spread in velocities a bit better, the median
clearly shows that most observed line-widths are very narrow. At an
approximate cold temperature of 15\,K, the thermal line-width of
$^{13}$CO(2--1) is 0.15\,km\,s$^{-1}$, which closely agree with the
observed median velocity dispersion of this nearby cloud. Hence,
R1near resembles a cold and mainly starless cloud with an almost
thermal velocity dispersion.

\subsubsection{The Coalsack cloud R2 comprising Tapia's Globule 2}

\begin{figure}[htb]
\includegraphics[width=90mm]{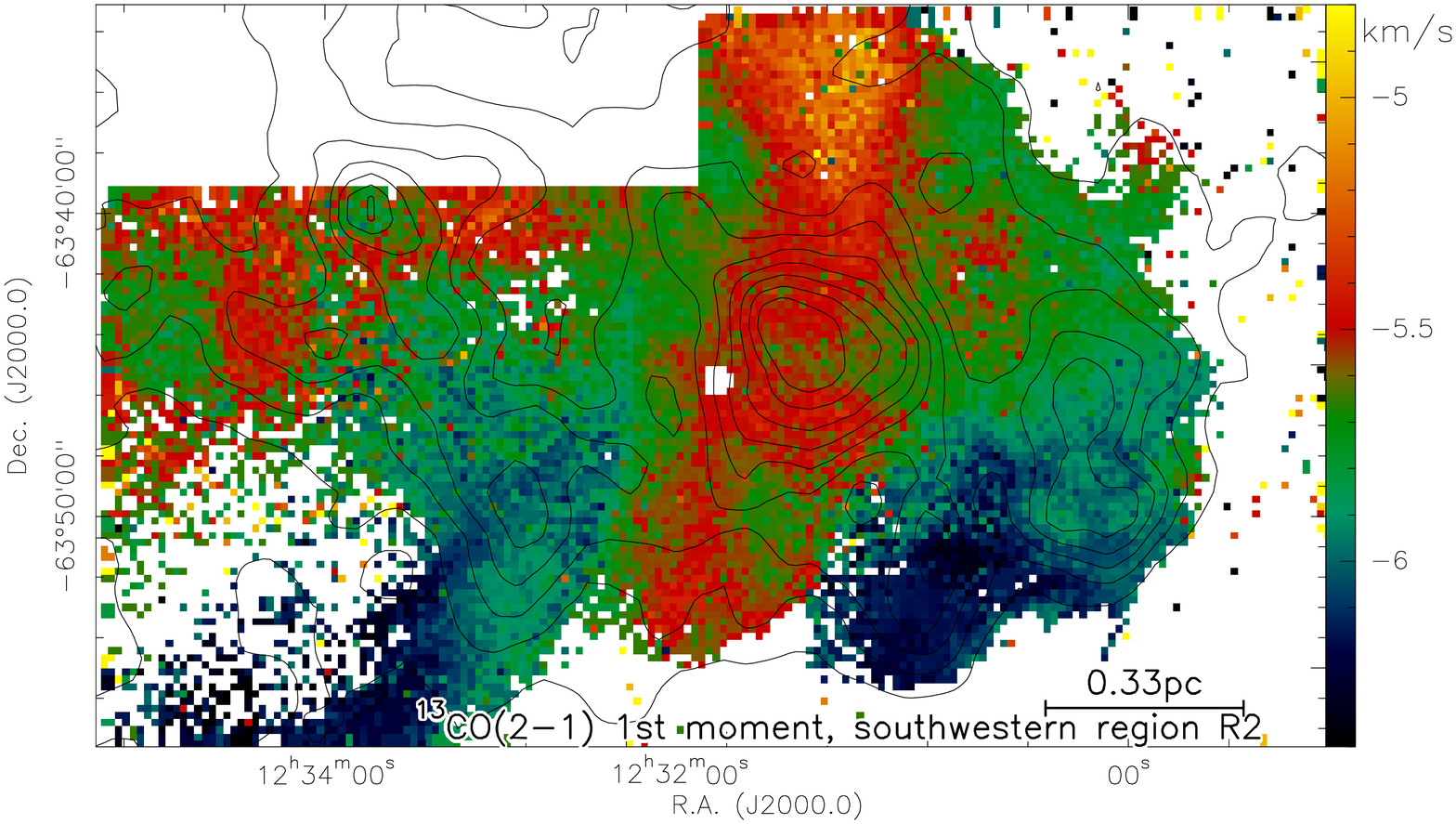}
\includegraphics[width=90mm]{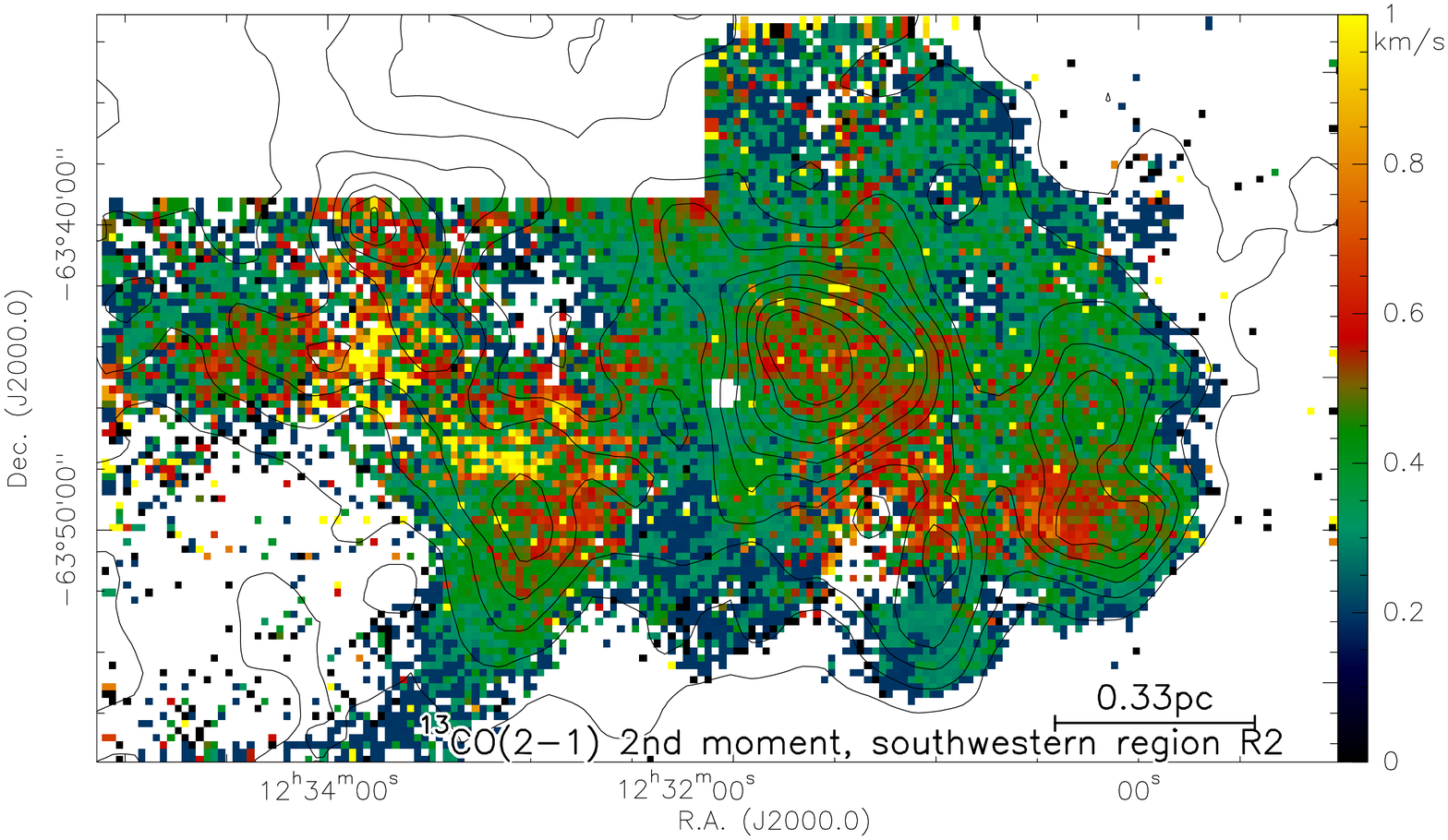}
\caption{Compilation of $^{13}$CO(2--1) 1st and 2nd moment maps toward
  the southwestern region R2 in the top and bottom panel,
  respectively.  The thick contours outline the extinction from
  \citet{kainulainen2009} in 1\,mag steps starting at an $A_v$ of
  2\,mags. Linear scale-bars are shown in the bottom-right corner of
  each panel.}
\label{moments_r2}
\end{figure}

Figure \ref{moments_r2} presents the 1st and 2nd moment maps
(intensity-weighted peak-velocities and velocity dispersion) for the
southwestern region R2, also known as ``Tapia's Globule 2''
\citep{tapia1973,lada2004,rathborne2009b}. The peak-velocity
distribution in this region does not exhibit as clear a velocity
gradients as seen in R1far and R1near (see previous sections), but we
find velocity gradients from the east and west of the main globule
toward the extinction peak. East and west of the main peak, the 1st
moment map reveals peak velocities $\sim$-5.8\,km\,s$^{-1}$, whereas
in a relatively broad north-south filament extending almost across the
central extinction peak, the peak velocities are found to be around
$-5.5$\,km\,s$^{-1}$.  Hence, in these $^{13}$CO(2--1) data we do not
recover the east-west velocity gradient previously discussed by
\citet{rathborne2009b} for the inner $6'\times 6'$ around the main
extinction peak.

The velocity dispersion of this region is also very narrow, spanning a
range mainly below 1\,km\,s$^{-1}$ (Fig.~\ref{moments_r2} bottom
panel).  In contrast to the R1 region where the average and mean
values of the velocity dispersions deviated from each other, which
indicatis a broader velocity spread, for R2, the average and mean
velocity dispersion are approximately the same with
$\sim$0.4\,km\,s$^{-1}$.  This can be interpreted as a sign of a
relatively uniform velocity dispersion over the region. While the
velocity dispersion at the edge of the map is around
0.2\,km\,s$^{-1}$, in the outskirts of the main extinction peak it
rises to values on the order of $\sim$0.4\,km\,s$^{-1}$, and toward
the extinction peak one finds values on the order of of
0.6\,km\,s$^{-1}$ or a bit higher.  While even the latter values are
still narrow compared to more active star-forming regions (e.g.,
\citealt{goldsmith2008}), they nevertheless are indicative of
nonthermal gas motions in the direction of the highest extinction. For
a more detailed discussion about the dynamical properties of the
region, see section \ref{dynamics}.

\section{Discussion}
\label{discussion}

\subsection{Fragmenting filaments}
\label{frag_fil}

The overall extent of the two clouds -- ``Nessie'' discussed in
\citet{jackson2010} and R1far presented here -- is similar with
$\sim$80\,pc and $\sim$88\,pc, respectively, and they are at similar
distances (both at $\sim$3.1\,kpc). However, there are also
considerable differences between the two clouds. First of all, the
length-to-width ratio is different because while ``Nessie'' is
extremely narrow with an approximate cylinder radius $r_{\rm{cyl}}$ of
only $\sim$0.5\,pc, the structure within R1far is less well defined
and hence it is also more difficult to accurately determine a radius.
However, based on the extinction maps, we roughly estimate an
approximate average radius $r_{\rm{cyl}}$ of $\sim 2-2.5$\,pc for
R1far, about four to five times larger than that of ``Nessie'',
derived from their MIPSGAL mid-infrared emission.  Furthermore,
\citet{jackson2010} report that star-formation activity is clearly
identified toward many of their clumps, while our analysis
reveals barely any star formation activity toward R1far. In addition to this,
as discussed in section \ref{results}, because the mass of R1far is
distributed over such a large spatial extent, it cannot be considered
as a ``typical'' very young high-mass star-forming region or IRDC.


How does this structure compare to a gravitationally bound isothermal
gaseous cylinder? Following \citet{jackson2010} as well as the
original work by \citet{chandrasekhar1953b}, \citet{nagasawa1987} and
\citet{inutsuka1992}, we try to estimate characteristic fragmentation
scales for different conditions within self-gravitating gas cylinders.
For an incompressible fluid, a characteristic length scale can be
defined as $\lambda_{\rm{frag}} \approx 11r_{\rm{cyl}}$, whereas for
an infinite isothermal gas cylinder, the relation is
$\lambda_{\rm{frag}} \approx 22H$ where $H$ is the isothermal scale
height given by $H = c_s(4\pi G\rho)^{-1/2}$ with $c_s$ the sound
speed, $G$ the gravitational constant and $\rho$ the central gas mass
density.

Depending on the sound speed $c_s$ we can calculate the characteristic
scale height of such a filament. With a thermal sound speed of
  the gas at 15\,K of $c_s\sim 0.23$\,km\,s$^{-1}$ and using
$n_{\rm{crit}}\sim 9\times 10^3$\,cm$^{-3}$ as a proxy for the
density, the characteristic scale-height is $H\approx 0.07$\,pc.
However, inspecting Fig.~\ref{moment2_r1}, we find that the region
rather exhibits an average velocity dispersion of
$\sim$1.26\,km\,s$^{-1}$, which we use as a proxy for the
full-width-half-maximum FWHM (see Sec.~\ref{r1far}). If one now
replaces the sound speed $c_s$ with the Gaussian velocity dispersion
$\sigma = 1/2.35 * \rm{FWHM}$, we obtain a higher scale-height of
$H\approx 0.16\,pc$. This implies that the approximate cylinder radius
$r_{\rm{cyl}}\sim 2.25$\,pc in any case exceeds the characteristic
scale height by more than an order of magnitude. Because the
characteristic length scale $\lambda_{\rm{frag}} \approx
11r_{\rm{cyl}}$ for an incompressible fluid requires $r_{\rm{cyl}}\ll H$
(e.g., \citealt{jackson2010}), the incompressible case can safely be
excluded for R1far.

Looking at the characteristic fragmentation scale for the infinite
isothermal gas cylinder $\lambda_{\rm{frag}} \approx 22H$, we derive
$\lambda_{\rm{frag}} \approx 1.46$\,pc or $\lambda_{\rm{frag}} \approx
3.5$\,pc for the sound speed $c_s$ or the Gaussian velocity
dispersion $\sigma$, respectively. Because our data are clearly not
described by a pure thermal line-width (Fig.~\ref{moment2_r1}), the
fragmentation scale $\lambda_{\rm{frag}} \approx 3.5$\,pc for a
Gaussian velocity dispersion consistent with our data appears the most
reasonable characteristic length scale we can derive in the framework
of this picture of a gravitationally bound isothermal gaseous
cylinder.

Comparing this latter $\lambda_{\rm{frag}} \approx 3.5$\,pc with our
derived median clump separation of $\sim$4\,pc (Sec.~\ref{r1far}), the
data are broadly consistent with a compressible gravitationally bound
isothermal gaseous cylinder, in particular if one considers the
observed spread in velocity dispersion values as well as an intrinsic
spread in the gas densities. However, they are inconsistent with
  an incompressible fluid.

  Another way to characterize self-gravitating cylinders is in the
  framework of maximum critical mass per unit length $M/l_{\rm{max}}$
  along the cylinder's axis
  \citep{ostriker1964,fiege2000a,fiege2000b}.  Above the maximum
  critical linear mass density, the cylinders would be doomed to
  collapse. Following again \citet{jackson2010} and using the mean
  $^{13}$CO(2--1) line-width of $\Delta v\approx 1.26$\,km\,s$^{-1}$,
  for turbulently supported filaments we find a maximum critical
  $M/l_{\rm{max}} = 84(\Delta
  v)^2$\,M$_{\odot}$pc$^{-1}=133$\,M$_{\odot}$pc$^{-1}$ (about a
  factor 10 higher than the maximum critical $M/l_{\rm{max}}\sim
  15$\,M$_{\odot}$pc$^{-1}$ for undisturbed thermal filaments,
  \citealt{ostriker1964,inutsuka1997}).  To compare that with our
  observations, we can obtain a rough estimate of the observed $M/l$ by
  dividing the total mass estimated from the $^{13}$CO data in
  Sec.~\ref{results} ($\sim$2600\,M$_{\odot}$) by the approximate
  length of the filament of $\sim$73\,pc in the $^{13}$CO data. This
  results in an approximate observed $M/l$ of only
  $\sim$36\,M$_{\odot}$pc$^{-1}$.  Although this is larger than the
  maximum critical linear mass density for thermal filaments (and also
  the measured $M/l$ ratio in, e.g., a thermal filament in Taurus,
  \citealt{schmalzl2010}), the observed mass-per-length scale is
  approximately a factor 4 below the maximum critical $M/l_{\rm{max}}$
  ratio for turbulently supported filaments based on the
  $^{13}$CO(2--1) line-width. While a lower than critical
  $M/l_{\rm{max}}$ does not necessarily imply stability against
  collapse, it is at least consistent with the Coalsack at large not
  (or only barely) forming stars at the moment.

  Comparing our results with those from the literature, we find that
  in the framework of isothermal gravitationally bound cylinders, we
  can distinguish for R1far between a compressible and an
  incompressible configuration, whereas this is not possible for the
  ``Nessie'' IRDC \citet{jackson2010}. Furthermore, the $M/l$ ratio
  for R1far with $\sim$36\,M$_{\odot}$pc$^{-1}$ is considerably lower
  than that for ``Nessie'' or Orion ($\sim$110\,M$_{\odot}$pc$^{-1}$
  and $\sim$385\,M$_{\odot}$pc$^{-1}$, respectively,
  \citealt{jackson2010}).  While one cannot directly compare the two
  latter values with the threshold we estimated above because the
  observed line-width in these clouds are also larger than in the
  Coalsack, nevertheless, this comparison confirms the expected
  evolution that starless clouds have low $M/l$ ratios that increase
  with time during the collapse and star-formation processes. However,
  not all low $M/l$ clouds have to evolve like that. They could also
  be transient structures that are dispersing again.

\subsection{Dynamical properties of R2/Tapia's Globule 2}
\label{dynamics}

\begin{figure}[htb]
\includegraphics[width=90mm]{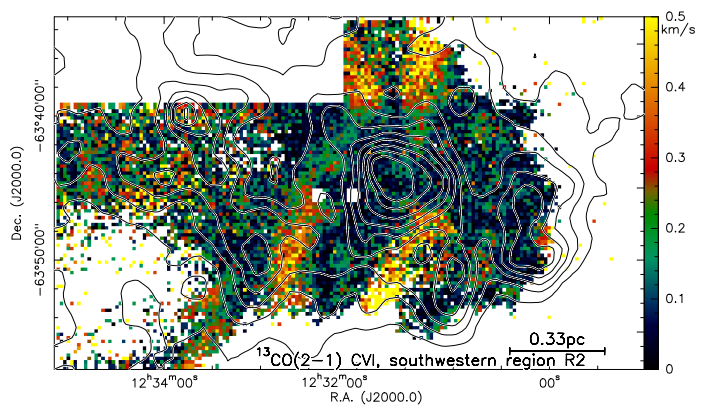}
\caption{$^{13}$CO(2--1) central velocity increment (CVI) in R.A.
  toward the southwestern region R2.  The thick contours outline the
  extinction from \citet{kainulainen2009} in 1\,mag steps starting at
  an $A_v$ of 2\,mags. A linear scale-bar is shown in the
  bottom-right corner.}
\label{cvi_r2}
\end{figure}

\citet{rathborne2009b} recently suggested that the ring-like structure
within this globule may have formed from the merging of two sub-sonic
flows. In contrast to \citet{rathborne2009b}, we do not find a
velocity gradient across the globule. This discrepancy likely arises
from different observational biases, e.g., we observed the region in
$^{13}$CO, whereas \citet{rathborne2009b} employed the rarer
isotopologue C$^{18}$O. Furthermore, their C$^{18}$O map covered a
much smaller area of the globule, and if one focuses on the central
few arcminutes of our data (Fig.~\ref{moments_r2}, top panel), one
could also get the impression of a velocity gradient.  Nevertheless,
the velocity dispersion increase toward the center is suggestive of
dynamical evolution within the core. To further investigate the
dynamics of the region, we produced a central-velocity-increment (CVI,
e.g., \citealt{hily-blant2008}) map of the region, which is a different
representation of the velocity changes over the observed region. Because
the velocity structure is more prominent in east-west direction, the
CVI map presented in Figure \ref{cvi_r2} studies the velocity changes
in the R.A.  direction with 10 pixel step sizes, corresponding to $\sim
143''$ or approximately five resolution elements (section \ref{obs}).
The observed peak-velocity changes are small toward the central peak
of the extinction map, however, we see four enhancements of the
CVI-values toward the eastern and western edges in the north and south
of the main extinction peak.  This is consistent with the fact that we
see a filamentary structure going from north to south through Globule
2 in the 1st moment map (Fig.~\ref{moments_r2}). Returning to the
merging subsonic flow picture by \citet{rathborne2009b}, our data do
not completely confirm this picture because we do not see a clear
east-west velocity gradient across as \citet{rathborne2009b} do.
However, the combined facts that we find clear CVI increases at the
Globule's edges as well as an increase of velocity dispersion toward
the extinction peak are strongly suggestive of a dynamically
active state of this dark globule within the large Coalsack complex.
From these data we cannot establish the exact reason for the dynamical
evolution, however, this Globule is an excellent candidate region for
future active collapse and star-formation processes.

\section{Conclusions}

Mapping approximately 1 square degree around the regions of the
highest extinction in the almost entirely starless Coalsack dark cloud
in $^{13}$CO(2--1) with APEX revealed several interesting properties:

\begin{itemize}

\item The Coalsack is not one coherent dark low-mass cloud at
  distances below 200\,pc and typical velocities of $\sim -4$ to
  0\,km\,s$^{-1}$, but we find a second velocity component at $\sim
  -30$\,km\,s$^{-1}$ corresponding to a second cloud at a distance of
  $\sim 3.1$\,kpc. This second more distant cloud dominates the
  dust-extinction signatures in the northeast of the Coalsack.

\item Mid-infrared MSX data reveal barely any star-formation activity
  toward any of the mapped clouds.

\item Although the total observed mass of the distant cloud with
  $\sim$ 2600\,M$_{\odot}$ is far larger than those of the two mapped
  nearby clouds (on the order of several solar masses), we cannot
  consider it as a potential high-mass star-forming region, because
  its mass is distributed over a spatial extent in east-west
  direction of $\sim$73\,pc. Hence, it is smoothly distributed over a
  large spatial area.

\item The more distant cloud resembles in shape a large and twisted
  filament. Analyzing this structure within the framework of an
  isothermal gravitationally bound cylinder, we find that the data are
  consistent with a compressible gaseous filament and inconsistent
  with an incompressible fluid. Furthermore, the characteristic mass
  per length $M/l$ is low with $\sim$36\,M$_{\odot}$pc$^{-1}$. Hence,
  this structure is potentially stable against gravitational
  collapse.

\item The nearby low-mass clouds both exhibit narrow velocity
  distributions with median values between 0.2 and
  0.4\,km\,s$^{-1}$, also indicative of almost no star-formation
  activity.

\item Only Tapia's Globule 2 appears different. While the peak
  velocity dispersion barely exceeds 0.6\,km\,s$^{-1}$, we
  nevertheless see a line-width increase from the map edge toward the
  extinction peak. Furthermore, an analysis of the
  central-velocity-increments reveals significant velocity changes at
  the Globule's edges. Both features are suggestive of dynamical
  action in that region, e.g., maybe early infall activity.

\end{itemize}

In summary, our data reveal different cloud components nearby as well
as far away. While all observed clouds are mostly bare of any
star-formation activity, we find filamentary structures consistent
with compressible self-gravitating gaseous cylinders as well as one
Globule in an apparent elevated dynamical state.

\begin{acknowledgements} 
  This research made use of data products from the Midcourse Space
  Experiment. Processing of the data was funded by the Ballistic
  Missile Defense Organization with additional support from NASA
  Office of Space Science. This research has also made use of the
  NASA/IPAC Infrared Science Archive, which is operated by the Jet
  Propulsion Laboratory, California Institute of Technology, under
  contract with the National Aeronautics and Space
  Administration.
\end{acknowledgements}


\end{document}